\documentclass[conference]{IEEEtran}

\IEEEoverridecommandlockouts

\usepackage{cite}
\usepackage{amsmath,amssymb,amsfonts}
\usepackage{graphicx}
\usepackage{textcomp}
\usepackage{xcolor}
\def\BibTeX{{\rm B\kern-.05em{\sc i\kern-.025em b}\kern-.08em
    T\kern-.1667em\lower.7ex\hbox{E}\kern-.125emX}}

\usepackage{booktabs}
\usepackage{tabularx}
\usepackage{caption}
\usepackage{subcaption}

\usepackage{algorithm}
\usepackage{algpseudocode}
\usepackage{comment}

\usepackage{hyperref}

\usepackage{graphicx}

\usepackage{enumitem}

\usepackage{cite}

\usepackage{tikz}
\usetikzlibrary{positioning,calc,arrows.meta,shapes.geometric,shapes.symbols}

\usepackage{pifont}

\newcommand{\bm}[1]{\boldsymbol{#1}}
\allowdisplaybreaks

\usepackage{tikz}
\usetikzlibrary{shapes.geometric,shapes.symbols,positioning,arrows.meta}
\usetikzlibrary{positioning,arrows.meta,decorations.pathreplacing,fit}
\tikzset{>=Latex}

\usepackage{enumitem}
\setlist{nosep,leftmargin=1.2em}

\begin{document}

\title{Fine-Tuning LLMs to Generate Economical and Reliable Actions
for the Power Grid
\thanks{This work was funded by NSF Grant 2130706 and ARO Grant W911NF2310266.}
}

\author{
\IEEEauthorblockN{
Mohamad Chehade and 
Hao Zhu
}
\IEEEauthorblockA{
Chandra Family Department of Electrical and Computer Engineering \\
The University of Texas at Austin, Austin, TX, USA \\
\{chehade, haozhu\}@utexas.edu
}
}

\maketitle

\begin{abstract}
Public Safety Power Shutoffs (PSPS) force rapid topology changes that can render standard operating points infeasible, requiring operators to quickly identify corrective transmission switching actions that reduce load shedding while maintaining acceptable voltage behavior. We present a verifiable, multi-stage adaptation pipeline that fine-tunes an instruction-tuned large language model (LLM) to generate \emph{open-only} corrective switching plans from compact PSPS scenario summaries under an explicit switching budget. First, supervised fine-tuning distills a DC-OPF MILP oracle into a constrained action grammar that enables reliable parsing and feasibility checks. Second, direct preference optimization refines the policy using AC-evaluated preference pairs ranked by a voltage-penalty metric, injecting voltage-awareness beyond DC imitation. Finally, best-of-$N$ selection provides an inference-time addition by choosing the best feasible candidate under the target metric. On IEEE 118-bus PSPS scenarios, fine-tuning substantially improves DC objective values versus zero-shot generation, reduces AC power-flow failure from 50\% to single digits, and improves voltage-penalty outcomes on the common-success set. Code and data-generation scripts are released to support reproducibility.
\end{abstract}



\section{Introduction}
\label{sec:intro}
Large language models (LLMs) have rapidly transitioned from research prototypes to deployable decision-support tools across diverse domains~\cite{mastropaolo2023copilot,zhang2025scientificllm,lai2024legalllm}. Their ability to transform unstructured descriptions into structured outputs makes them attractive for operational environments where decisions are time-sensitive and consequences are high. Power system control rooms are a compelling setting: operators manage complex contingencies, coordinate actions across many assets, and balance reliability, economics, and compliance under tight time constraints~\cite{lin2013mcrworkload}. Unlike traditional decision-support tools that require specialized inputs or rigid interfaces, LLMs enable operators to interact through natural language while producing machine-readable recommendations (e.g., structured action lists) that can be verified before execution~\cite{kalami2025llmgrid,sanguinetti2024conversational}.

However, foundation LLMs lack domain-specific knowledge of power system physics, operational constraints, and grid safety requirements. Training grid-specific LLMs \emph{from scratch} is impractical: modern LLMs succeed through pre-training on trillions of tokens spanning diverse domains~\cite{kaplan2020scaling}, while grid operations data are orders of magnitude smaller and specialized. A practical alternative is to \emph{adapt} a strong instruction-tuned model via targeted fine-tuning so that it can (i) read a compact, structured description of a grid scenario and (ii) output actions in a constrained grammar that can be checked for feasibility.

In this work, we study a concrete and operationally motivated task: corrective, \emph{open-only} transmission switching during Public Safety Power Shutoffs (PSPS), which are corrective de-energization actions used by utilities to reduce wildfire ignition risk during extreme weather conditions~\cite{pge2024psps}. When PSPS forces lines out of service, operators must rapidly determine whether opening additional elements can mitigate overloads, reduce load shedding, and improve operating conditions while respecting switching budgets and operational rules. Computing optimal actions with mixed-integer optimization can be expensive under time pressure, especially when considering nonlinear AC constraints~\cite{fisher2008ots,hedman2009otsca,haag2024opsdcac}. Our goal is to amortize this optimization effort into training, then produce high-quality switching recommendations at inference time using structured scenario summaries and a verifiable action grammar.

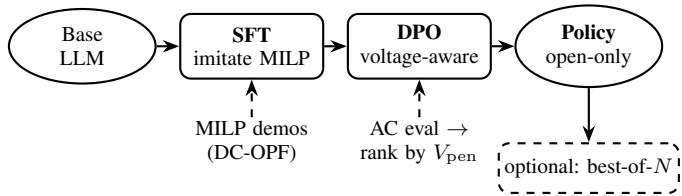
\begin{figure}[t]
\centering
\begin{tikzpicture}[
  font=\footnotesize,
  box/.style={draw, rounded corners, thick, align=center,
              minimum width=1.90cm, minimum height=0.86cm, inner sep=2.8pt},
  end/.style={draw, thick, ellipse, align=center,
              minimum width=1.95cm, minimum height=0.86cm, inner sep=2.8pt},
  opt/.style={draw, rounded corners, thick, dashed, align=center,
              minimum width=1.95cm, minimum height=0.66cm, inner sep=2.2pt},
  arrow/.style={-{Stealth[length=2.2mm]}, thick},
  sup/.style={-{Stealth[length=2.2mm]}, thick, dashed},
  lab/.style={align=center, inner sep=1.0pt}
]

\node[end] (base) {Base\\LLM};
\node[box, right=0.30cm of base] (sft) {\textbf{SFT}\\imitate MILP};
\node[box, right=0.30cm of sft] (dpo) {\textbf{DPO}\\voltage-aware};
\node[end, right=0.30cm of dpo] (pol) {\textbf{Policy}\\open-only};

\draw[arrow] (base) -- (sft);
\draw[arrow] (sft) -- (dpo);
\draw[arrow] (dpo) -- (pol);

\node[lab, below=0.50cm of sft, text width=1.90cm] (oracle) {MILP demos\\(DC-OPF)};
\draw[sup] (oracle.north) -- (sft.south);

\node[lab, below=0.50cm of dpo, text width=2.25cm] (prefs) {AC eval $\rightarrow$\\rank by $V_{\mathrm{pen}}$};
\draw[sup] (prefs.north) -- (dpo.south);

\node[opt, below=0.72cm of pol] (bon) {optional: best-of-$N$};
\draw[arrow] (pol.south) -- (bon.north);

\end{tikzpicture}
\vspace{-2mm}
\caption{Multi-stage adaptation pipeline for PSPS corrective switching.}
\label{fig:methodology_pipeline}
\vspace{-2mm}
\end{figure}

Figure~\ref{fig:methodology_pipeline} illustrates the pipeline we adopt. Starting from an instruction-tuned base model, supervised fine-tuning (SFT) trains the LLM to imitate MILP-derived open-only switching decisions under DC constraints. We then apply direct preference optimization (DPO) using ranked responses derived from AC voltage-quality evaluation, producing a voltage-aware policy that more reliably prioritizes actions with fewer voltage violations.

This design follows a standard alignment pattern for instruction-tuned LLMs: imitation learning first, followed by preference-based refinement~\cite{ouyang2022instructgpt,rafailov2024direct}. In our setting, the supervised stage anchors the policy to an optimization oracle, while the preference stage injects AC voltage-awareness that is difficult to encode directly in DC training. The resulting model functions as a candidate-plan generator whose outputs can be parsed, verified, and evaluated with existing grid-analysis tools. Our contributions are:
\begin{itemize}[leftmargin=*,topsep=4pt,itemsep=2pt]
  \item We formulate PSPS-aware open-only switching with switching budgets and corridor structure using a DC-OPF MILP oracle (Section~\ref{sec:psps}).
  \item We design a structured scenario representation and action grammar that enables an instruction-tuned LLM to emit switching plans that are straightforward to parse and verify (Section~\ref{sec:sft}).
  \item We introduce a voltage-aware preference refinement stage based on DPO, using AC-derived voltage-quality preferences to align the model beyond DC imitation (Section~\ref{sec:dpo}).
  \item We evaluate economic performance, AC feasibility, and voltage quality, including comparisons to a neural baseline and training-curve reporting for reproducibility (Section~\ref{sec:experiments}).
\end{itemize}

Finally, we discuss practical considerations such as feasibility checks, training/inference costs, and deployment constraints (Section~\ref{sec:experiments}). We view this as a step toward verifiable, operator-facing LLM assistants that interface with existing grid analysis pipelines rather than replacing them.

\section{PSPS Switching Problem}
\label{sec:psps}
Public Safety Power Shutoffs (PSPS) are preventive and corrective de-energization actions taken by utilities to reduce wildfire ignition risk during extreme weather conditions~\cite{rhodes2021ops,kody2022pspsinvest}. When a PSPS event forces a subset of transmission lines out of service, system operators must determine whether additional \emph{corrective open-only} switching actions can improve reliability and reduce load shedding. We formulate a DC optimal power flow (DC-OPF) model that explicitly incorporates PSPS constraints and pose an open-only decision-making problem in which operators may proactively open a limited number of additional transmission elements to mitigate system stress.

\paragraph{Network model and PSPS constraints.}
Consider a power system with bus set $\mathcal{B}$ ($|\mathcal{B}| = n_b$), transmission line set $\mathcal{E}$ ($|\mathcal{E}| = n_\ell$), and generator set $\mathcal{G} \subseteq \mathcal{B}$. Each line $e$ has reactance $x_e > 0$ and rating $S^{\max}_e > 0$ (MW). Buses have demand $P^{\mathrm{d}}_i \ge 0$, generators have capacity $P^{\min}_g \le P_g \le P^{\max}_g$ and cost $c_g \ge 0$ (\$/MW), and load shedding $P^{\mathrm{s}}_i \ge 0$ is penalized at rate $\gamma$ (\$/MW). A PSPS event is encoded by availability mask $\bm{\xi} \in \{0,1\}^{n_\ell}$, where $\xi_e = 0$ forces line $e$ open. Operator decisions $z_e \in \{0,1\}$ determine effective status $g_e = \xi_e z_e$. We use standard DC power flow with bus-branch incidence $A$ and susceptance $B_\ell = \mathrm{diag}(1/x_e)$.

\paragraph{Open-only switching with a line budget.}
Operators may open up to $K_\ell$ additional PSPS-available lines. The open-only switching problem is:
\begin{subequations}
\label{eq:milp-line}
\begin{align}
  \min_{\substack{\bm{P}_g,\bm{P}^{\mathrm{s}},\bm{\theta},\bm{P}_\ell,\\ \bm{z} \in \{0,1\}^{n_\ell}}} \quad
  & \sum_{g \in \mathcal{G}} c_g P_g \;+\; \gamma \sum_{i \in \mathcal{B}} P^{\mathrm{s}}_i \\
  \text{s.t.} \quad
  & P^{\min}_g \le P_g \le P^{\max}_g,\ \forall g \in \mathcal{G}, \\
  & P^{\mathrm{s}}_i \ge 0,\ \forall i \in \mathcal{B}, \\
  & \bm{P}_\ell = B_\ell A^\top \bm{\theta},\ \theta_r = 0, \\
  & A \bm{P}_\ell = \bm{P}_g - (\bm{P}^{\mathrm{d}} - \bm{P}^{\mathrm{s}}), \\
  & -S^{\max}_e\, \xi_e z_e \le P_{\ell,e} \le S^{\max}_e\, \xi_e z_e,\ \forall e \in \mathcal{E}, \label{eq:milp-thermal} \\
  & z_e \in \{0,1\},\ z_e = 0\ \text{if}\ \xi_e = 0,\ \forall e \in \mathcal{E}, \label{eq:milp-fix} \\
  & \sum_{e \in \mathcal{E}} (1 - z_e)\,\xi_e \le K_\ell. \label{eq:milp-budget}
\end{align}
\end{subequations}
Constraint~\eqref{eq:milp-fix} enforces that PSPS-forced outages remain open; \eqref{eq:milp-budget} limits operator-induced opens to $K_\ell$ available lines. This is structurally related to optimal transmission switching~\cite{fisher2008ots,hedman2009otsca} but restricted to open-only actions.

\paragraph{Corridor-constrained open-only switching.}
Switching decisions may be constrained to transmission corridors $\mathcal{S}$—geographically grouped lines~\cite{moreira2024wildfireddp,kody2022pspsinvest}. Binary variables $y_S \in \{0,1\}$ indicate whether corridor $S$ is activated for switching. Corridor and line decisions are coupled by:
\begin{subequations}
\label{eq:corridor}
\begin{align}
  z_e &\ge 1 - y_S, \quad \forall S \in \mathcal{S},\ \forall e \in S, \label{eq:corridor-couple} \\
  \sum_{S \in \mathcal{S}} y_S &\le K_S,\quad y_S \in \{0,1\},\ \forall S \in \mathcal{S}, \label{eq:corridor-budget}
\end{align}
\end{subequations}
When $y_S = 0$, \eqref{eq:corridor-couple} prevents operator opens in corridor $S$; when $y_S = 1$, line decisions remain free. Constraint~\eqref{eq:corridor-budget} limits activated corridors to $K_S$.

\paragraph{Role in the pipeline.}
We use the DC-OPF MILP above as an \emph{offline oracle} to generate labeled switching plans for supervised adaptation. Voltage quality and AC feasibility are evaluated separately using AC power flow in the experimental section, enabling us to train on DC structure while assessing voltage-critical performance.

\section{Supervised Fine-Tuning (SFT) LLMs}
\label{sec:sft}
Solving the mixed-integer linear program (MILP) in~\eqref{eq:milp-line} for every PSPS scenario can be computationally expensive, particularly when evaluating large numbers of contingencies or when operators require rapid what-if analysis. Similar scalability challenges are well documented for optimal transmission switching (OTS) formulations, motivating learning-assisted and proxy-based approaches that approximate an optimizer while preserving most of the economic benefit~\cite{pineda2024learningOTS,bugaje2023rttswitch}. We therefore train a large language model (LLM) to \emph{imitate} an optimization oracle, amortizing the cost of offline MILP solves into a single supervised fine-tuning (SFT) phase. After SFT, the model generates candidate switching plans from compact scenario summaries, which are then parsed and verified before evaluation or deployment. This section describes the oracle, the input--output representation, and the SFT protocol.

\paragraph{Ground-Truth Oracle.}
Given PSPS mask $\bm{\xi}$ and budget $K_\ell$, we solve the DC-OPF MILP (\ref{eq:milp-line}) to obtain optimal operator-induced opens
\begin{equation}
  \mathcal{T}(\bm{\xi}) \;\triangleq\; \bigl\{\,e \in \mathcal{E} : \xi_e = 1,\; z^\star_e = 0\,\bigr\}.
  \label{eq:oracle-toggles}
\end{equation}
For corridor-constrained problems, we solve the variant with ~\eqref{eq:corridor}.

\paragraph{Scenario Representation.}
Each training example corresponds to a PSPS scenario with mask $\bm{\xi}$ and budgets $K_\ell$, $K_S$. We provide a structured textual summary including case dimensions, the number of PSPS-forced opens, and a corridor breakdown listing member lines with their forced/eligible status and applicable budgets. This abstracts away numerical parameters while preserving topological structure.

\paragraph{Action Grammar.}
The target encodes $\mathcal{T}(\bm{\xi})$ using: \texttt{open(Sk:LINE)} for corridor-associated lines (e.g., \texttt{open(S6:135)}), \texttt{open(LINE)} for non-corridor lines, and \texttt{do\_nothing} if $\mathcal{T}(\bm{\xi}) = \emptyset$. We sort actions lexicographically to form a canonical string, e.g.,
\[
  \texttt{open(S3:21); open(S7:98); open(131)}.
\]

\paragraph{Dataset Formatting.}
Supervised pairs use a chat format: system prompt defining the task and grammar, user message with scenario JSON, and assistant message with the canonical action string, following standard instruction-tuning practice~\cite{ouyang2022instructgpt}.

\paragraph{Fine-Tuning Objective and Protocol.}
We initialize from an instruction-tuned base LLM and perform supervised fine-tuning on the training partition to learn a policy $\pi_{\mathrm{SFT}}$ that maps scenario summaries to action strings. The objective is the standard conditional language-modeling loss over the assistant tokens. Let $s_k$ denote the scenario summary and $a_k$ the canonical action string for example $k$. We optimize parameters $\phi$ by minimizing
\begin{equation}
  \mathcal{L}_{\mathrm{SFT}}(\phi)
  \;=\;
  -\sum_{k} \log p_\phi\bigl(a_k \,\big|\, s_k,\ \text{system prompt}\bigr),
  \label{eq:sft-loss}
\end{equation}
where the sum ranges over training examples. This mirrors the supervised stage used in instruction-following pipelines prior to preference optimization~\cite{ouyang2022instructgpt}. The resulting model $\pi_{\mathrm{SFT}}$ is then used for inference and (optionally) subsequent preference-based refinement.

\paragraph{Parsing and Verification.}
At inference, we parse outputs and verify: (i) grammar validity, (ii) all opened lines are PSPS-available ($\xi_e = 1$), and (iii) budget adherence. Invalid outputs are flagged, ensuring the LLM functions as a candidate generator rather than an unverified controller.

\subsection{Improving Voltage  Safety via DPO (SFT-Time)}
\label{sec:dpo}
The supervised policy $\pi_{\mathrm{SFT}}$ imitates a DC open-only oracle, but DC imitation alone does not explicitly optimize voltage quality under nonlinear AC physics. To better align the policy with voltage safety objectives, we introduce a refinement stage based on \emph{direct preference optimization} (DPO)~\cite{rafailov2024direct}. DPO is an reinforcement learning-free objective that trains a policy from pairwise preferences $(y^+,y^-)$, encouraging the refined policy to assign higher probability to actions that yield better voltage outcomes under AC evaluation.

\paragraph{Voltage-Penalty Metric.}
For candidate plan $y$, we parse into topology $\bm{z}(y)$ and solve AC power flow to obtain bus voltage magnitudes $\{|V_i(y)|\}$. We define a deadband $v_{\mathrm{db}}$ around nominal voltage and penalize violations outside $[1-v_{\mathrm{db}},\,1+v_{\mathrm{db}}]$:
\begin{equation}
  V_{\mathrm{pen}}(y)
  \;\triangleq\;
  \kappa \sum_{i\in\mathcal{B}} \Bigl(\max\bigl\{\,\bigl||V_i(y)|-1\bigr|-v_{\mathrm{db}},\;0\,\bigr\}\Bigr)^p,
  \label{eq:vpen}
\end{equation}
where $p\in\{1,2\}$ selects aggregation norm and $\kappa>0$ scales the penalty. Non-convergent AC solutions receive penalty $V_{\mathrm{fail}}$ (large constant).

\paragraph{Preference Pair Construction.}
For each scenario $x$, we sample $N_{\mathrm{pref}}$ candidates from $\pi_{\mathrm{SFT}}(\cdot\mid x)$, discard malformed or budget-violating plans, evaluate $V_{\mathrm{pen}}(y)$ via AC power flow, and select $(y^+, y^-)$ pairs where
\begin{equation}
    V_{\mathrm{pen}}(y^-) - V_{\mathrm{pen}}(y^+) \ge \Delta_{\mathrm{pref}},
    \label{eq:pref-gap}
\end{equation}
yielding preference dataset $\mathcal{D}_{\mathrm{volt}}=\{(x_i,y_i^+,y_i^-)\}_{i=1}^{M}$ where $y^+$ is preferred for voltage quality.

\paragraph{DPO Objective.}
Let $\pi_{\phi}$ denote the trainable policy initialized from $\pi_{\mathrm{SFT}}$, with reference $\pi_{\mathrm{ref}}= \pi_{\mathrm{SFT}}$. For triple $(x,y^+,y^-)$, define
\begin{equation}
  \Delta_{\phi}(x,y^+,y^-)
  \;\triangleq\;
  \log \pi_{\phi}(y^+ \mid x)-\log \pi_{\phi}(y^- \mid x),
\end{equation}
and analogously $\Delta_{\mathrm{ref}}$. The DPO loss is
\begin{equation}
  \mathcal{L}_{\mathrm{DPO}}(\phi)
  \;=\;
  -\!\!\sum_{(x,y^+,y^-)\in\mathcal{D}_{\mathrm{volt}}}
  \log \sigma\!\Bigl(\beta_{\mathrm{DPO}}\bigl[\Delta_{\phi}-\Delta_{\mathrm{ref}}\bigr]\Bigr),
  \label{eq:dpo-loss}
\end{equation}
where $\beta_{\mathrm{DPO}}>0$ controls preference strength~\cite{rafailov2024direct}. Minimizing this produces $\pi_{\mathrm{DPO}}$, which increases likelihood of low-penalty plans while anchored to the SFT reference.

\paragraph{Scope and Practicality.}
Preference construction uses AC power flow \emph{offline} to label candidates with voltage-quality information. At inference time, the refined model can be used either in single-shot mode or combined with best-of-$N$ reranking (Section~\ref{sec:bon}), where $V_{\mathrm{pen}}$ can serve as a selection criterion when AC evaluation is available.


\subsection{Improving Voltage Safety via Best-of-$N$ (Inference-Time)}
\label{sec:bon}
Even after supervised fine-tuning and preference alignment, a single stochastic decode can produce a suboptimal or malformed plan. We therefore use \emph{best-of-$N$} inference as a lightweight, training-free mechanism to improve solution quality by sampling multiple candidate plans and selecting the best under a task metric. Best-of-$N$ is widely used in reasoning and structured prediction tasks to exploit sampling diversity, often with strong gains at moderate $N$~\cite{wang2023selfconsistency}.

Given a scenario summary $x$ and policy $\pi$ (e.g., $\pi_{\mathrm{SFT}}$ or $\pi_{\mathrm{DPO}}$), we draw $N$ independent candidates:
\begin{equation}
  y^{(j)} \sim \pi(\cdot \mid x),\qquad j=1,\dots,N.
  \label{eq:bon-sample}
\end{equation}
Each candidate is verified through staged checks: grammar parsing, PSPS/budget constraint validation, DC feasibility evaluation, and optional AC power flow for voltage scoring. Let $\mathcal{Y}_{\mathrm{valid}}(x)$ denote candidates passing verification. If empty, we fall back to a safe default (e.g., doing nothing). We then select the plan minimizing a scalar score:
\begin{equation}
  \hat{y}(x)
  \;\in\;
  \arg\min_{y \in \mathcal{Y}_{\mathrm{valid}}(x)} \; \mathrm{Score}(x,y).
  \label{eq:bon-select}
\end{equation}
Primary choices are: (i) DC economic score $J_{\mathrm{DC}}(x,y)$, or (ii) AC voltage penalty $V_{\mathrm{pen}}(y)$ from~\eqref{eq:vpen}. When both matter, we use scalarization $\mathrm{Score}(x,y)=J_{\mathrm{DC}}(x,y) + \lambda\, V_{\mathrm{pen}}(y)$ with $\lambda \ge 0$ set by operational priorities.

Best-of-$N$ increases compute linearly in $N$. We use moderate $N$ with early stopping when target scores are met. Sampling is embarrassingly parallel, and $N$ can be adjusted to trade off compute for quality. We quantify costs and inference times in Section~\ref{sec:experiments}.


\section{Experimental Results}
\label{sec:experiments}
Experiments use the IEEE 118-bus test system from MATPOWER~\cite{matpower} ($n_b = 118$, $n_\ell = 186$, $n_g = 54$). We define $|\mathcal{S}| = 9$ transmission corridors by grouping geographically proximate lines (8--20 lines per corridor). PSPS events are simulated by selecting corridor lines to force open, producing availability mask $\bm{\xi}$ for each scenario. Unless stated otherwise, operators may open at most $K_\ell = 3$ additional PSPS-available lines.

\paragraph{Datasets.}
For SFT, we use {200 PSPS scenarios} split between training and held-out testing. For DPO, we construct {440 preference pairs} $(x,y^+,y^-)$ by sampling candidates from $\pi_{\mathrm{SFT}}$ and ranking by AC voltage penalty $V_{\mathrm{pen}}$ (Section~\ref{sec:dpo}). Data-generation scripts and splits are in the released codebase.

\paragraph{Implementation and Models.}
Power flow and optimization use MATLAB R2024b with YALMIP~\cite{lofberg2004yalmip}. We adapt an instruction-tuned LLM, \texttt{ft:gpt-4.1-mini-2025-04-14}, via the OpenAI fine-tuning API. SFT trains for 3 epochs (batch size 1, LR multiplier 2) on 453{,}384 training tokens. DPO initializes from $\pi_{\mathrm{SFT}}$ and trains for 2 epochs (batch size 8) with $\beta_{\mathrm{DPO}}=0.1$ on 1{,}611{,}736 tokens. For the evaluation of the voltage penalties, we use $v_{\mathrm{db}}=0$, $p=1$, $\kappa=1$ for \eqref{eq:vpen}. Our pipeline is backend-agnostic and can be applied to any instruction-tuned LLM deployed either via API or locally hosted models.

\paragraph{Compared Policies.}
We compare: (i) zero-shot base LLM, (ii) $\pi_{\mathrm{SFT}}$, (iii) $\pi_{\mathrm{DPO}}$, and (iv) a neural network baseline: a fully-connected MLP with one hidden layer of width 512 (ReLU), trained on the same supervised dataset to predict corrective opens under identical budget and feasibility constraints.

\paragraph{Code Release.}
Full code is available on GitHub: \href{https://github.com/MFHChehade/LLM-Grid-Actions}{MFHChehade/LLM-Grid-Actions}.

\subsection{Training Curves for the Fine-Tuning Process}
\begin{figure}[t]
  \centering
  \begin{minipage}{0.48\linewidth}
    \centering
    \includegraphics[width=\linewidth]{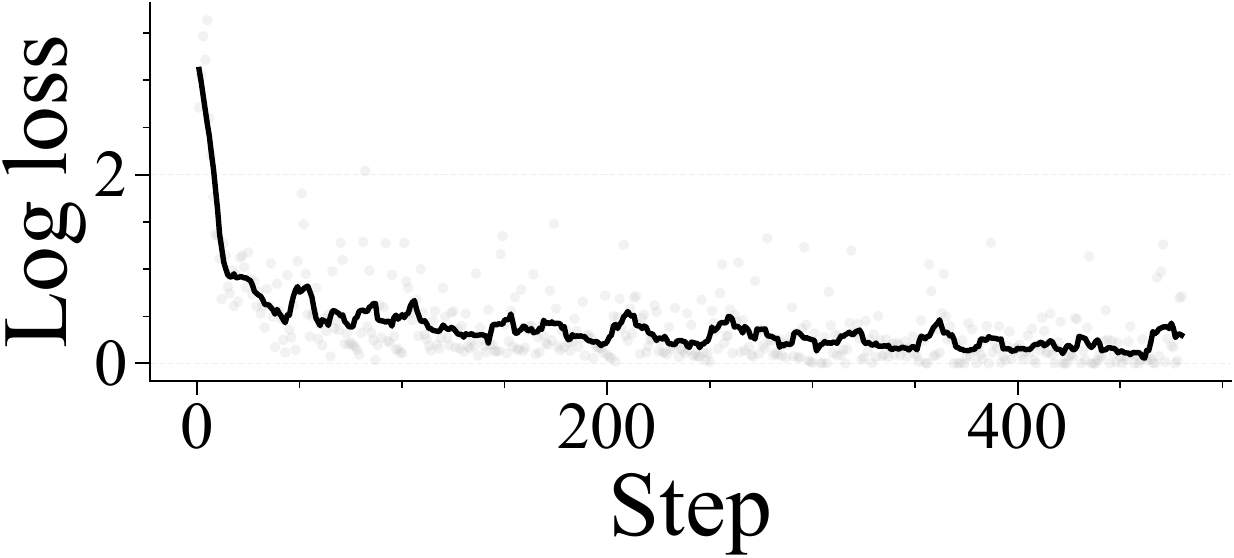}
    \vspace{-1mm}
    {\small (a) SFT log loss.}
  \end{minipage}
  \hfill
  \begin{minipage}{0.48\linewidth}
    \centering
    \includegraphics[width=\linewidth]{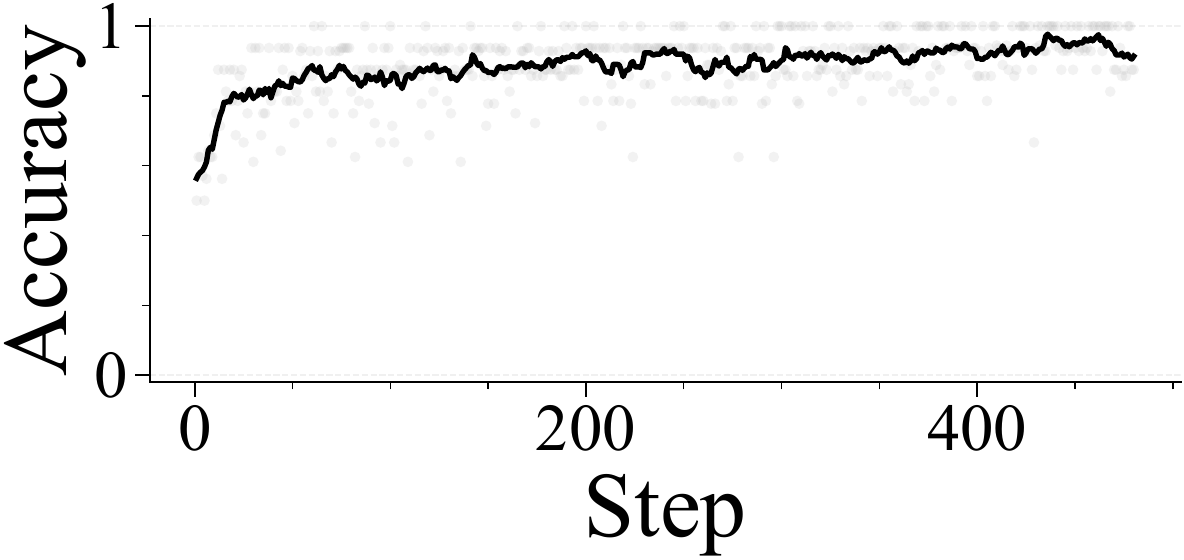}
    \vspace{-1mm}
    {\small (b) SFT token accuracy.}
  \end{minipage}

  \vspace{1mm}

  \begin{minipage}{0.48\linewidth}
    \centering
    \includegraphics[width=\linewidth]{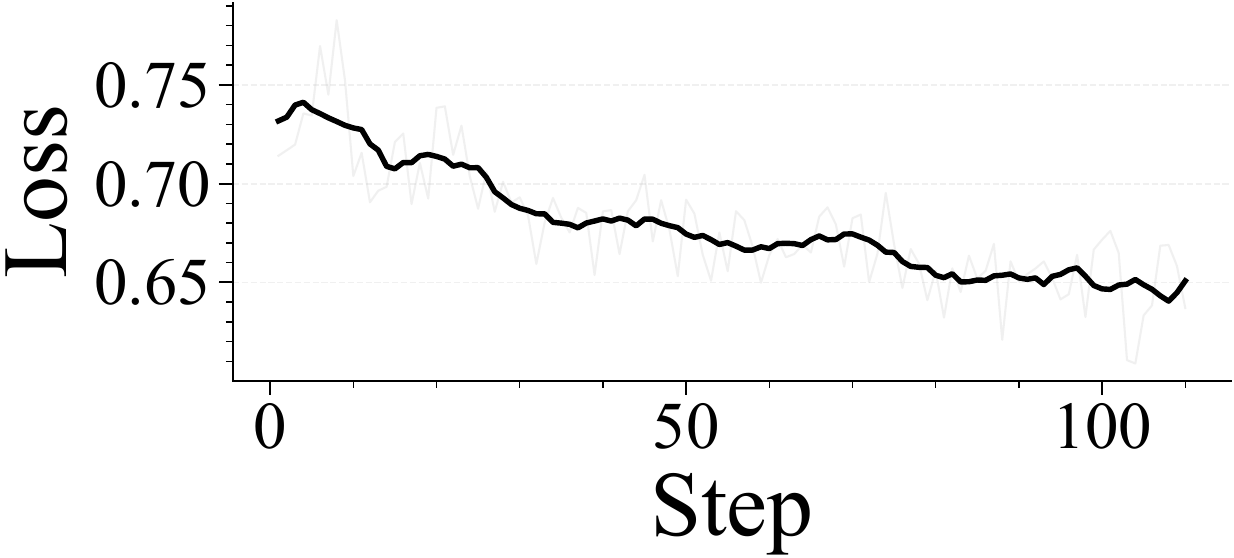}
    \vspace{-1mm}
    {\small (c) DPO loss.}
  \end{minipage}
  \hfill
  \begin{minipage}{0.48\linewidth}
    \centering
    \includegraphics[width=\linewidth]{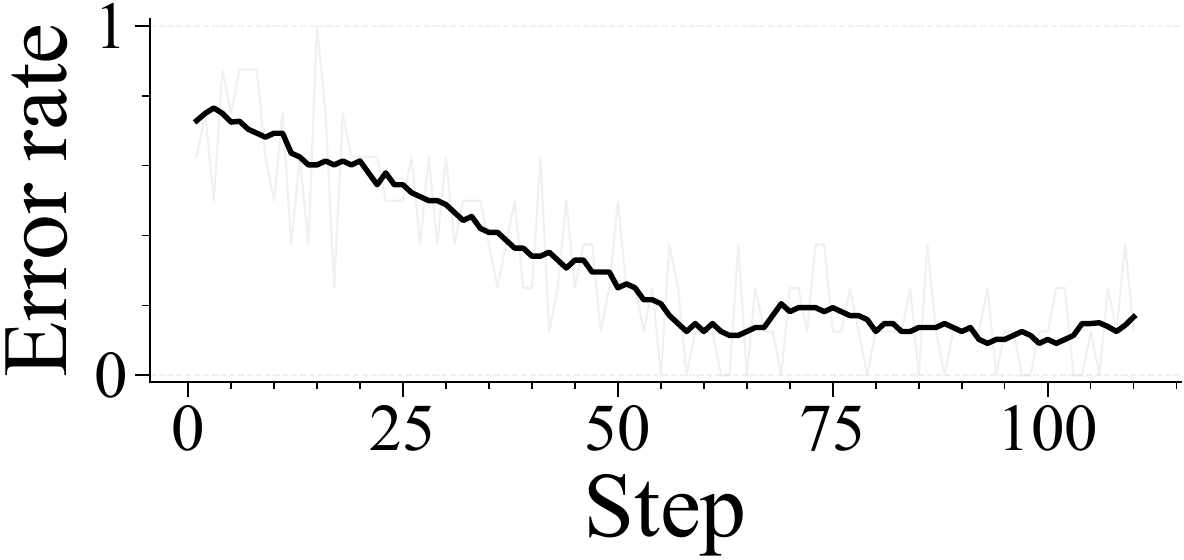}
    \vspace{-1mm}
    {\small (d) DPO error rate.}
  \end{minipage}

  \caption{Training curves from the fine-tuning jobs (SFT and DPO) show the convergence.}
  \label{fig:train_curves}
\end{figure}

Figures~\ref{fig:train_curves}(a--b) show that the SFT process is well-behaved: log loss drops sharply early then decreases gradually, while token accuracy rapidly increases and stabilizes. This pattern reflects the model learning the output grammar then refining scenario-to-action mappings. Figure~\ref{fig:train_curves}(c--d) shows stable DPO optimization with decreasing loss and preference error rate, indicating successful separation of preferred vs.\ dispreferred plans. No divergence occurs under the chosen $\beta_{\mathrm{DPO}}$ and dataset size.

\subsection{DC Objective}

\begin{figure}[t]
  \centering
  \includegraphics[width=\linewidth]{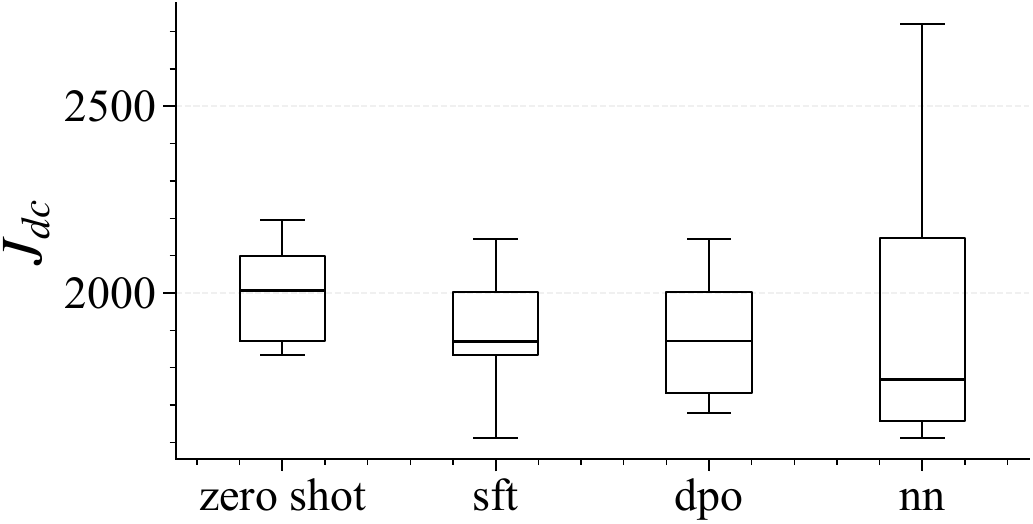}
  \caption{Distribution of DC objective $J_{\mathrm{DC}}$ across all four compared policies (zero-shot, SFT, DPO, and NN).}
  \label{fig:box_Jdc_models_only}
\end{figure}

Figure~\ref{fig:box_Jdc_models_only} reports the box plots for the $J_{\mathrm{DC}}$ distributions. Both SFT and DPO shift downward relative to zero-shot, distilling the oracle signal into lower-cost decisions. SFT and DPO medians are close, showing voltage-aware preference refinement preserves DC performance. The NN baseline achieves low median $J_{\mathrm{DC}}$ but exhibits wider dispersion and heavier upper tail, indicating occasional high-cost decisions.

\subsection{AC Feasibility and Voltage Quality}

\begin{figure}[t]
  \centering
  \includegraphics[width=\linewidth]{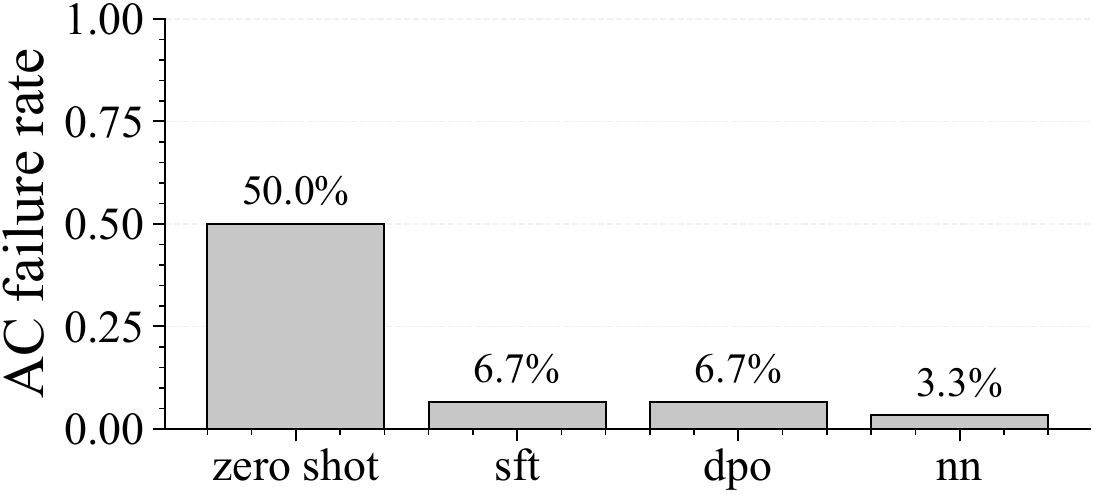}
  \caption{AC power-flow failure rate across compared policies. Fine-tuned policies drastically reduce AC failures relative to the zero-shot baseline.}
  \label{fig:ac_fail_rate}
\end{figure}

\begin{figure}[t]
  \centering
  \includegraphics[width=\linewidth]{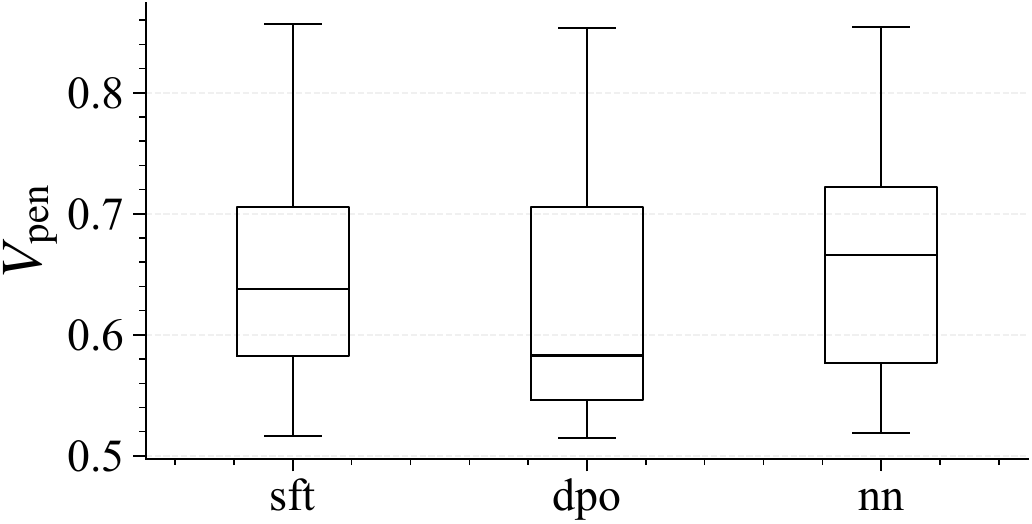}
  \caption{Voltage penalty $V_{\mathrm{pen}}$ distribution on the \emph{common-success set}, i.e., scenarios where all compared policies (SFT, DPO, NN) have achieved AC convergence. This ensures an apples-to-apples comparison of voltage quality.}
  \label{fig:box_Vpen_common_success}
\end{figure}

Figure~\ref{fig:ac_fail_rate} shows zero-shot fails AC power flow in half of scenarios, while SFT and DPO reduce failures to a small fraction. This indicates DC oracle imitation plus constrained grammar substantially improves physical plausibility. The NN baseline achieves the lowest failure rate.

We evaluate $V_{\mathrm{pen}}$ on the \emph{common-success set} (Figure~\ref{fig:box_Vpen_common_success})—scenarios where SFT, DPO, and NN all converge. DPO achieves lower median $V_{\mathrm{pen}}$ than SFT and NN, consistent with preference refinement improving voltage outcomes beyond DC imitation. However, the upper tail persists across policies, suggesting some scenarios remain challenging for voltage regulation and motivating richer preference data as future work.

\section{Conclusion}
\label{sec:conclusion}
We developed a verifiable fine-tuning pipeline that adapts a foundation LLM into a corrective switching assistant for PSPS events. Supervised fine-tuning distills MILP-derived DC-optimal \emph{open-only} actions into a constrained, parseable grammar that enables systematic feasibility checks. Direct preference optimization then injects AC voltage-awareness by learning from preference pairs ranked using a voltage-penalty metric, improving voltage quality among AC-feasible cases. Best-of-$N$ selection complements training by trading inference compute for improved candidate quality. Empirically on IEEE 118-bus PSPS scenarios, the fine-tuned policies outperform zero-shot generation in economic objective, dramatically reduce AC infeasibility, and yield improved voltage-penalty distributions on the common-success set, while remaining compatible with standard power-flow verification. Future work will investigate multi-task fine-tuning of a single foundation model so it can support diverse power-grid applications under a unified, verifiable decision framework.



\bibliographystyle{IEEEtran}
\bibliography{ref}

\end{document}